\documentclass[preprint,showpacs,preprintnumbers,amsmath,amssymb,graphicx]{revtex4}
\usepackage{amsmath,amsfonts,amssymb}
\usepackage{epsfig}
\def\bit{\begin{itemize}}
\def\eit{\end{itemize}}
\def\ben{\begin{enumerate}}
\def\een{\end{enumerate}}
\def\bed{\begin{description}}
\def\eed{\end{description}}

\def\lsim{\raise0.3ex\hbox{$<$\kern-0.75em\raise-1.1ex\hbox{$\sim$}}}
\def\gsim{\raise0.3ex\hbox{$>$\kern-0.75em\raise-1.1ex\hbox{$\sim$}}}

\let\jnfont=\rm
\def\NPB#1,{{\jnfont Nucl.\ Phys.\ B}{\bf #1},}
\def\PLB#1,{{\jnfont Phys.\ Lett.\ B}{\bf #1},}
\def\EPJC#1,{{\jnfont Eur.\ Phys.\ Jour.\ C}{\bf #1},}
\def\PRD#1,{{\jnfont Phys.\ Rev.\ D}{\bf #1},}
\def\PRL#1,{{\jnfont Phys.\ Rev.\ Lett.\ }{\bf #1},}
\def\MPLA#1,{{\jnfont Mod.\ Phys.\ Lett.\ A}{\bf #1},}
\def\JPG#1,{{\jnfont J.\ Phys.\ G}{\bf #1},}
\def\CTP#1,{{\jnfont Commun.\ Theor.\ Phys.\ }{\bf #1},}
\def\JHEP#1,{{\jnfont JHEP }{\bf #1},}
\def\NPPS#1,{{\jnfont Nucl.\ Phys.\ Proc.\ Suppl.\ }{\bf #1},}

\def\beq{\begin{equation}}
\def\eeq{\end{equation}}
\def\bea{\begin{eqnarray}}
\def\eea{\end{eqnarray}}
\newcommand{\ba}{\begin{array}}
\newcommand{\ea}{\end{array}}

\def\endash{\hbox{--}}
\def\Det{\mathop{\rm Det}}

\begin{document}

\title{Light dark matter in NMSSM and implication on Higgs phenomenology}

\author{ Junjie Cao$^1$, Ken-ichi Hikasa$^2$, Wenyu Wang$^3$, Jin Min Yang$^4$  }

\affiliation{
$^1$ Department of Physics, Henan Normal university, Xinxiang 453007, China\\
$^2$ Department of Physics, Tohoku University, Sendai 980-8578, Japan \\
$^3$ Institute of Theoretical Physics, College of Applied Science,
     Beijing University of Technology, Beijing 100124, China\\
$^4$ Institute of Theoretical Physics, Academia Sinica,
              Beijing 100190, China ~ \vspace{0.5cm} }

\begin{abstract}
For the experimental search of neutralino dark matter, it is
important to know its allowed mass and scattering cross section with
the nucleon. In order to figure out how light a neutralino dark
matter can be predicted in low energy supersymmetry, we scan over
the parameter space of the NMSSM (next-to-minimal supersymmetric
model), assuming all the relevant soft mass parameters to be below
TeV scale. We find that in the parameter space allowed by current
experiments the neutralino dark matter can be as light as a few GeV
and its scattering rate off the nucleon can reach the
sensitivity of XENON100 and CoGeNT. As a result, a
sizable parameter space is excluded by the current
XENON100 and CoGeNT data (the plausible CoGeNT dark matter signal
can also be explained).
The future 6000 kg-days exposure of
XENON100 will further explore (but cannot completely cover) the
remained parameter space.  Moreover, we find that in such a light
dark matter scenario a light CP-even or CP-odd Higgs boson must be
present to satisfy the measured dark matter relic density.
Consequently, the SM-like Higgs boson $h_{SM}$ may decay
predominantly into a pair of light Higgs bosons or a pair of
neutralinos so that the conventional decays like $h_{SM} \to
\gamma \gamma$ is much suppressed.
\end{abstract}
\maketitle

{\em Introduction:} Experiments for the underground direct detection
of cold dark matter $\tilde{\chi}$ have recently made significant
progress. While the null observation of $\tilde{\chi}$ in the CDMS
and XENON100 experiments has set rather tight upper limits on the
spin-independent (SI) cross section of $\tilde{\chi}$-nucleon
scattering \cite{CDMSII, XENON100}, the CoGeNT experiment
\cite{CoGeNT} reported an excess which cannot be explained by any
known background sources but seems to be consistent with the signal
of a light $\tilde{\chi}$ with mass around 10 GeV and scattering
rate around $10^{-40}$ cm$^2$. Intriguingly, this
range of mass and scattering rate is compatible with the dark matter
explanation for both the DAMA/LIBRA data and the preliminary CRESST
data \cite{Hooper}. So far, due to the inconsistency of the
CoGeNT result with the CDMS or XENON result, it is premature to draw
any definite conclusion about the existence or nonexistence of a light
$\tilde{\chi}$. However, considering much effort is being paid
on the search of a light $\tilde{\chi}$ in experiments,
it is theoretically important to check the
possible new physics prediction for a light
$\tilde{\chi}$ and examine
its related phenomenology (such as the Higgs boson search) at the LHC.
In this work we will focus on low energy supersymmetry, where the
lightest neutralino $\tilde{\chi}_1^0$ serves as the dark matter
candidate, and perform an intensive study of the light
$\tilde{\chi}_1^0$ scenario.

The most popular model for low energy supersymmetry is the minimal
supersymmetric standard model (MSSM). In this model, we find from
our scan that the neutralino $\tilde{\chi}_1^0$ must be heavier
than about 28 GeV. The main reason for the absence of a lighter
$\tilde{\chi}_1^0$ is its dominant annihilation channel is
$\tilde{\chi}_1^0\tilde{\chi}_1^0 \to b\bar{b}$ through
$s$-channel exchange of the pseudoscalar Higgs boson ($A$) and the
measured dark matter relic density requires $m_A \sim
(90\endash100)$ GeV and $\tan \beta \sim 50$, which is in conflict
with the constraints from the LEP and $B$ physics experiments
\cite{MSSM-light,NMSSM-light,Belanger}. Here we emphasize that for
the above region the effects of the charged Higgs on $B \to X_s
\gamma$ are unacceptably large, and even with a fine tuning of the
contributions from the stop/chargino diagrams, such large effects
cannot be reduced to an acceptable level. Our results for the MSSM
are in agreement with \cite{Belanger}, but differ from
\cite{newest} in which the considered constraints, such as the
invisible $Z$-decay and the productions of neutralinos or Higgs
bosons at LEP II,  are weaker than in our study. Our conclusion
also differs from \cite{Bottino} because we used more accurate
formula in calculating the process $B \to X_s \gamma$
\cite{Cao-bsg}. Since the neutralino dark matter in the MSSM
cannot be so light as suggested by the CoGeNT data (albeit not
corroborated by XENON100 or CDMS), here we do not present
our MSSM results in detail.

Another popular model for low energy supersymmetry is the
next-to-minimal supersymmetric standard model (NMSSM) \cite{NMSSM}
which extends the MSSM by adding one gauge singlet chiral
superfield $\hat{S}$. This model is well motivated because it
provides a solution to the $\mu$-problem and the little hierarchy
problem the MSSM suffers from. For this model we perform an
intensive scan over its parameter space by assuming all the relevant
soft mass parameters  below TeV scale and considering various
experimental constraints. We find that in this model
the neutralino dark matter
can be as light as a few GeV and its spin-independent scattering
cross section with the nucleon can reach the sensitivity of CoGeNT
and XENON100.

We emphasize that our study is not restricted to explain any
special experimental results like the CoGeNT or DAMA/LIBRA data.
Instead, we aim to investigate the characteristics of a light
$\tilde{\chi}_1^0$, such as its lower mass bound and scattering
rate off the nucleon, and also examine the related Higgs
phenomenology at the LHC.
\vspace*{0.5cm}

{\em The NMSSM:} We start our analysis by recapitulating some basics of the NMSSM.
Its superpotential and the associated
soft-breaking terms in the Higgs sector are given by \cite{NMSSM}
\begin{align}
\mathbf{W}&=\lambda \hat{S} \hat{H}_u \hat{H}_d + \frac{1}{3}\kappa \hat{S}^3,\\
 V_{\it soft} &= {m^2_{H_d}} |H_d|^2 + {m^2_{H_u}} |H_u|^2
+{m^2_S}|S|^2 \nonumber \\
 & + (\lambda A_{\lambda} H_u H_d S + h.c.)+\left(\frac{\kappa}{3} A_{\kappa} S^3 +
h.c.\right), \label{higgspot}
\end{align}
where $H_d$, $H_u$ and $S$ denote scalar components of the superfield
$\hat{H}_d$, $\hat{H}_u$ and $\hat{S}$, respectively. After using the
minimization condition of the Higgs potential, this sector is described
by three dimensionless parameters ($\tan \beta$, $\lambda$, $\kappa$)
and three dimensionful
parameters ($\mu$, $A_\lambda$, $A_\kappa$). Due to the imposed $Z_3$ symmetry,
the superpotential does not contain dimensionful parameters and thus
all dimensionful parameters are generated by the soft-breaking masses
which, as required by the electroweak symmetry breaking,
should be naturally below TeV scale \cite{NMSSM}.
Other free parameters are the same as in the MSSM,
i.e., the soft masses for sfermions and gauginos as well as the trilinear soft
couplings.

Due to the presence of $\hat{S}$, the NMSSM predicts five
neutralinos, three CP-even Higgs bosons ($h_{1,2,3}$) and two CP-odd
Higgs bosons ($a_{1,2}$) \cite{NMSSM}. In general, the neutralino
mass eigenstates are the mixture of the MSSM neutralino fields and
the singlino field which is the fermion component of $\hat{S}$; the
CP-even (odd) Higgs mass eigenstates are similarly the mixture of
the CP-even (odd) MSSM Higgs fields and the real (imaginary) part of
the scalar component of $\hat{S}$. An important feature of the NMSSM
is that one of the CP-even (odd) Higgs bosons may be singlet-like
and thus can be very light \cite{NMSSM}. This feature is
particularly useful for light $\tilde{\chi}_1^0$ scenario
since it opens up new important annihilation channels for
$\tilde{\chi}_1^0$, i.e., either into a pair of $h_1$ (or $a_1$) or
into a pair of fermions via $s$-channel exchange of $h_1$ (or
$a_1$) \cite{Belanger,light-anni,Cao-nMSSM}. For the former case,
$\tilde{\chi}_1^0$ must be heavier than $h_1$ ($a_1$); while for the
latter case, due to the very weak couplings of
 $h_1$ ($a_1$) with $\tilde{\chi}_1^0$ and with the SM fermions,
a resonance enhancement (i.e.  $m_{h_1}$ or  $m_{a_1}$ must be close
to $2m_{\tilde{\chi}_1^0}$) is needed to accelerate the
annihilation. So a light $\tilde{\chi}_1^0$ should be necessarily
accompanied by a light $h_1$ or $a_1$ to provide the required dark
matter relic density.

Now we discuss how to get  a light $h_1$ or $a_1$ in the NMSSM. A
light $a_1$ can be easily obtained when the theory approaches to
the U(1)$_R$ or U(1)$_{\rm PQ}$ symmetry limit, which can be realized
by setting the product $\kappa A_\kappa$ to be negatively small
\cite{NMSSM}. In contrast, a light $h_1$ can not be obtained
so easily, but, as shown below, it can still be achieved by somewhat
subtle cancellation via tuning the value of $A_\kappa$. We note
that for any theory with multiple Higgs fields, the existence of a
massless Higgs boson implies the vanishing of the determinant of
its squared mass matrix and vice versa. For the NMSSM, at tree
level the parameter $A_\kappa$ only enters the mass term of the
singlet Higgs bosons and thus the determinant ($\Det{\cal{M}}^2$) of
the mass matrix for the CP-even Higgs bosons depends on $A_\kappa$
linearly \cite{NMSSM}. When other relevant parameters are fixed,
one can then obtain a light $h_1$ by varying $A_\kappa$ around the
value $\tilde{A}_\kappa$ which is the solution to the equation
$\Det{\cal{M}}^2=0$. In practice, one must include the important
radiative corrections to the Higgs mass matrix, which will
complicate the dependence of ${\cal{M}}^2$ on $A_\kappa$. However,
we checked that the linear dependence is approximately maintained
by choosing the other relevant parameters at the SUSY scale, and
one can solve the equation iteratively to get the solution
$\tilde{A}_\kappa$.
\vspace*{0.5cm}

{\em Numerical scan and results: } In order to study light
$\tilde{\chi}_1^0$ scenario in the NMSSM, we scan randomly over the
parameters in the neutralino and Higgs sectors by requiring $ 0 \leq
\lambda, \kappa \leq 0.7$, $1 \leq \tan \beta \leq 60$, $ 0 \leq
\mu, A_\lambda, M_2 \leq 1 {\rm~TeV} $ and $ 0 \leq M_1 \leq 100
{\rm~GeV} $.  Here the ranges of $\lambda$, $\kappa$ and $\tan
\beta$ are determined by the perturbativity of the theory
\cite{NMSSM}, the ranges of $\mu$ and $A_\lambda$ are suggested by
the electroweak symmetry breaking, and the narrow range of bino mass
$M_1$ is chosen to facilitate a light $\tilde{\chi}_1^0$. Since the
gluino mass and the soft parameters in the squark sector affect
little on the properties of $\tilde{\chi}_1^0$, we set all of them
to be 1~TeV. As for the soft slepton parameters, since they
influence the muon anomalous magnetic moment $a_\mu$ which in turn
can limit the important parameter $\tan \beta$, we assign them a
common scale $m_{\tilde{l}}$ and vary it below TeV scale.

Since a light $\tilde{\chi}_1^0$ is very likely to be accompanied
by a light $h_1$ or $a_1$, we perform two independent scans
aiming at a light $h_1$ and a light $a_1$ respectively.
For the light $h_1$ case, we vary $A_\kappa$ around $\tilde{A}_\kappa$
which is obtained by solving the equation $\Det{\cal{M}}^2=0$;
while for the light $a_1$ case, we simply vary $A_\kappa$ in the range
[$-200$, 0] GeV.

In our scans, we require all dimensionful parameters in the Higgs
potential like $m_{H_u}$ and $m_{H_d}$ below TeV scale and keep the
parameter points which yield $m_{\tilde{\chi}_1^0} \leq 20$ GeV.
The constraints considered in our scan are the following
\cite{Cao-JHEP}: (1) We require $\tilde{\chi}^0_1$ to account for
the dark matter relic density $0.105 < \Omega h^2 < 0.119$; (2) We
require the NMSSM contribution to explain the deviation of the muon
$a_\mu$, i.e., $a_\mu^{\rm exp} - a_\mu^{\rm SM} = ( 25.5 \pm 8.0 )
\times 10^{-10}$, at $2 \sigma$ level; (3) The LEP-I bound on the
invisible $Z$-decay, $\Gamma(Z \to \tilde{\chi}^0_1
\tilde{\chi}^0_1) < 1.76$ MeV, and the LEP-II upper bound on
$\sigma(e^+e^- \to \tilde{\chi}^0_1 \tilde{\chi}^0_i)$, which is $5
\times 10^{-2}~{\rm pb}$ for $i>1$, as well as the lower mass bounds
on sparticles from direct searches at LEP and the Tevatron; (4) The
constraints from the direct search for Higgs bosons at LEP-II,
including the decay modes $h \to h_1 h_1, a_1 a_1 \to 4 f$, which
limit all possible channels for the production of the Higgs bosons;
(5) The constraints from $B$ physics observables such as $B \to X_s
\gamma$, $B_s \to \mu^+\mu^-$, $B_d \to X_s \mu^+ \mu^-$, $B^+ \to
\tau^+ \nu$, $\Upsilon \to \gamma a_1 $, the $a_1$--$\eta_b$ mixing
and the mass difference $\Delta M_d$ and $\Delta M_s$; (6) The
constraints from the precision electroweak observables such as
$\rho_{\rm lept}$, $\sin^2 \theta_{\rm eff}^{\rm lept}$, $m_W$ and
$R_b$; (7) The constraints from the decay $\Upsilon(1S) \to \gamma
h_1 \to \gamma (\pi^+ \pi^-, K^+ K^-)$,  $\Upsilon(nS) \to \gamma
h_1 \to \gamma \mu^+ \mu^-$ ($n=1,2,3$) and the Tevatron search for
a light Higgs boson via $4 \mu$ and $2 \mu 2 \tau$ signals
\cite{Dark-Higgs}. The constraints (1--5) have been encoded in the
package NMSSMTools \cite{NMSSMTools}. We use this package in our
calculation and extend it by adding the constraints (6, 7). As
pointed out in \cite{Dark-Higgs}, the constraint (7) is important
for a light Higgs boson.

%%%%fig.1 %%%%%%%%%%%%%%%%%%%%%%%%%%%%%%%%%%%%%%%%%%%
\begin{figure}[htbp]
\includegraphics[width=15cm]{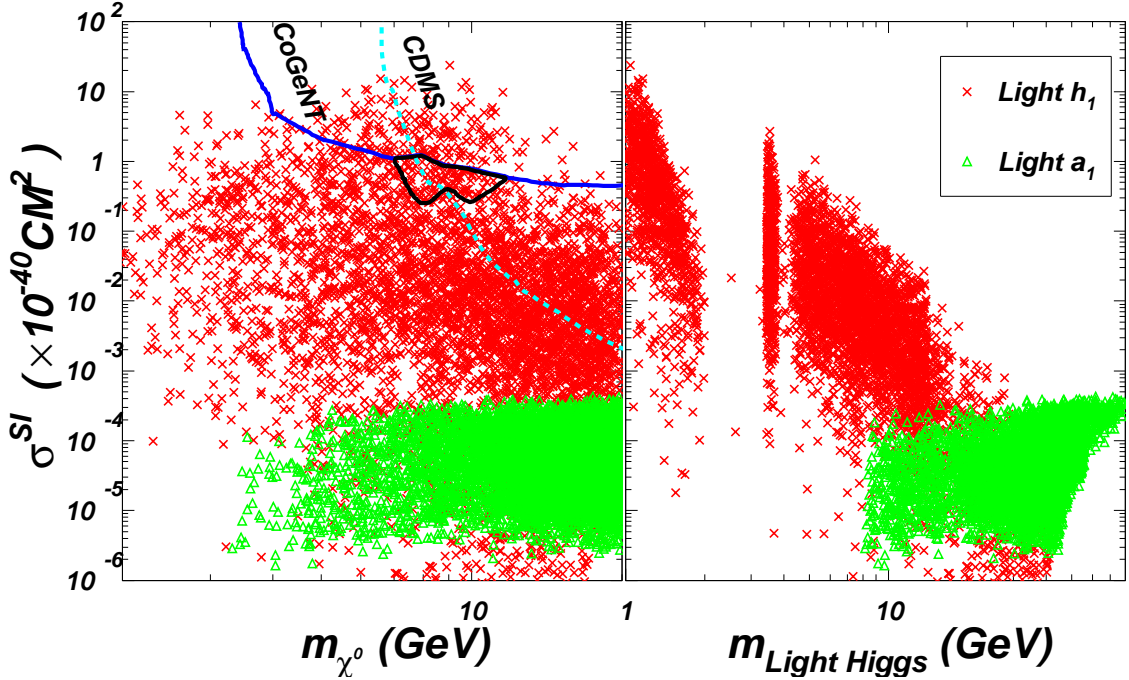}
\vspace{-0.5cm} \caption{The scatter plots of the parameter samples
which survive all constraints. The samples denoted by `$\times$' (red),
which are simultaneously projected on the $\sigma^{\rm SI}$-$m_\chi$ plane
in the left frame and the $\sigma^{\rm SI}$-$m_{h_1}$ plane in the right
frame, are characterized by a light $h_1$; while
the samples denoted by `$\triangle$' (green),
which are simultaneously projected on the $\sigma^{\rm SI}$-$m_\chi$ plane
in the left frame and the $\sigma^{\rm SI}$-$m_{a_1}$ plane in the right
frame, are characterized by a light $a_1$.
The curves are the limits from CoGeNT \cite{CoGeNT}, CDMS \cite{CDMSII}
and XENON100 \cite{XENON100}, while the contour is the CoGeNT-favored
region \cite{CoGeNT}. The future XENON100 (6000 kg-days exposure) sensitivity
is also plotted \cite{XENON-6000}.} \label{fig1}
\end{figure}
%%%%%%%%%%%%%%%%%%%%%%%%%%%%%%%%%%%%%%%%%%%%%%%%%%%%%%%

In Fig.~\ref{fig1} we display the surviving parameter samples, which
are simultaneously projected on the $\sigma^{\rm SI}$-$m_\chi$ plane
in the left frame and the $\sigma^{\rm SI}$-$m_{h_1}$ or
$\sigma^{\rm SI}$-$m_{a_1}$ plane in the right frame.
In our calculation we use the
formula in \cite{susy-dm-review} for the scattering rate and choose
$f_{T_u}^{(p)} = 0.023$, $f_{T_d}^{(p)} = 0.034$, $f_{T_u}^{(n)} =
0.019$, $f_{T_d}^{(n)} = 0.041$ and $f_{T_s}^{(p)} = f_{T_s}^{(n)} =
0.020$ as input. Note that, motivated by recent lattice simulation
\cite{lattice}, we take a very small value of $f_{T_s}$ and
consequently our estimation of the rate is rather conservative.

The left frame of Fig.~\ref{fig1} clearly shows that in the NMSSM
the neutralino dark matter can be as light as several GeV, while the
right frame of Fig.~\ref{fig1} shows that such a light neutralino is
accompanied by either a light $h_1$ or a light $a_1$. So the
surviving samples were classified into two sets, characterized
respectively by a light $h_1$  and a light $a_1$. In both the
light-$h_1$ case and the light-$a_1$ case the surviving samples give
a bino-dominant $\tilde{\chi}_1^0$ and correspondingly a small $M_1$
($\lsim 30$ GeV). The allowed regions for other parameters are
listed in Table \ref{table1}. We see that the value of $\mu$ is not
so large (below 300 GeV in light-$h_1$ case and 350 GeV in
light-$a_1$ case) because a low $\mu$ can enhance the coupling of
$\tilde{\chi}_1$ to $h_1$ and $a_1$ to get the correct relic
density.  Also we see a moderately loose bound on $m_{a_1}$ (we
checked that among the surviving samples about $2\%$ have $m_{a_1}
\geq 60$ GeV). This is due to a possibly large $\tan \beta$, which
can enhance the couplings of $a_1$ to the SM fermions so that
$m_{a_1}$ may deviate from $2 m_{\tilde{\chi}_1^0} $ significantly.
In the light-$h_1$ case, about $75\%$ of the surviving samples are
found to satisfy $m_{h_1} \leq m_{\tilde{\chi}_1^0}$, in which the
annihilation mode $\tilde \chi_1^0\tilde \chi_1^0\to h_1 h_1 $ plays
a crucial role in getting the required dark matter relic density. In
contrast, in the light-$a_1$ case most surviving samples are found
to satisfy $m_{a_1} \geq m_{\tilde{\chi}_1^0}$, in which the
dominant annihilation channel of dark matter is $\tilde
\chi_1^0\tilde \chi_1^0\to a_1^* \to f\bar{f}$.

%%%Table 1 %%%%%%%%%%%%%%%%%%%%
\begin{table}[t]
\begin{center}
\caption{The allowed parameter ranges for the light-$h_1$ case (first
row) and light-$a_1$ case (second row). The dimensionful parameters
 $ \mu $, $A_\kappa$ and $m_{h_1,a_1}$ are in unit of GeV.
\label{table1}}
\begin{tabular}{|c|c|c|c|c|c|}
\hline $\lambda$ & $\kappa$  & $\tan \beta$ & $ \mu $    &
$A_\kappa$ & $m_{h_1,a_1}$   \\ \hline
0.15--0.7 &0--0.5& 1--7& 130--300 &$-600$--0&
                                      $1 \lsim m_{h_1} \lsim 45$ \\ \hline
0.15--0.5 &0.1--0.7& 8--60&150--350&$-80$--0
                                      & $8 \lsim m_{a_1} \lsim 80$
 \\ \hline
\end{tabular}
\end{center}
\end{table}
%%%%%%%%%%%%%%%%%%%%%%%%%%%%%%%%%%%%%%%%%%

From Fig.~\ref{fig1} we see that the scattering rate of the light
dark matter can reach the sensitivity of XENON100 and, consequently,
a sizable parameter space is excluded by the XENON100 (2011) data
\cite{XENON100}. The future XENON100 experiment (6000 kg-days
exposure) \cite{XENON-6000} can further explore (but cannot
completely cover) the remained parameter space. Note that in the
light-$h_1$ case the scattering rate can be large enough to reach
the sensitivity of CoGeNT and can cover the CoGeNT-favored region.
The underlying reason is that the $\chi$-nucleon scattering can
proceed through the $t$-channel exchange of the CP-even Higgs
bosons, which can be enhanced by a factor $1/m_{h_1}^4$ for a light
$h_1$ \cite{light-anni}; while a light $a_1$ can not give such an
enhancement because the CP-odd Higgs bosons do not contribute to the
scattering in this way.  We noticed that the studies in
\cite{NMSSM-light,Das-light} claimed that the NMSSM is unable to
 explain the CoGeNT data because they did not consider the
light-$h_1$ case.

Note that the light-$h_1$ samples
are separated into three regions of $m_{h_1}$, as shown in the right
frame of Fig.~\ref{fig1}.
This is due to the {\it{combined}}
constraints from the dark matter relic density and the processes
$\Upsilon(1S) \to \gamma h_1$ (with $h_1 \to \pi^+ \pi^-, K^+ K^-,
\mu^+ \mu^-$), $B_d \to X_s \mu^+ \mu^-$ and $p\bar{p} \to h_{SM} \to
4\mu ({\rm or~} 2 \mu 2 \tau)$ at the Tevatron, which tightly limited the
couplings and mass of $h_1$.  For example, since in this case
$h_1$ mainly acts as the product of dark matter annihilation,
its coupling to $\tilde{\chi}_1^0$ must be
moderately large to get the correct relic density, while its
couplings to the SM fermions and $h_{SM}$ must be small to
suppress the rates of the processes mentioned above \cite{Dark-Higgs}.
The situation is different for the light-$a_1$ case,
where the relic density and the
LEP search for Higgs boson require $m_{a_1} > 8 {\rm GeV}$.
For such a 'heavy' $a_1$ the above mentioned low energy processes
give no stringent constraints and thus the light-$a_1$ samples
are not separated into different regions of $m_{a_1}$, as
shown in the right frame of Fig.~\ref{fig1}.

{\em Implication on Higgs physics: } In the NMSSM the light
$\tilde{\chi}_1^0$ scenario may predict rather peculiar Higgs
phenomenology due to the presence of the light particles. Among the
predicted Higgs bosons, the SM-like Higgs boson $h_{\rm SM}$, defined as
the CP-even Higgs boson with largest couplings to $Z^0$ pair, will
be the most important one to be searched at the LHC since it mainly
responsible for electroweak symmetry breaking. So we focus on
$h_{\rm SM}$ in our following discussion.
%%%%fig.2 %%%%%%%%%%%%%%%%%%%%%%%%%%%%%%%%%%%%%%%%%%%
\begin{figure}[htb]
\includegraphics[width=14.0cm]{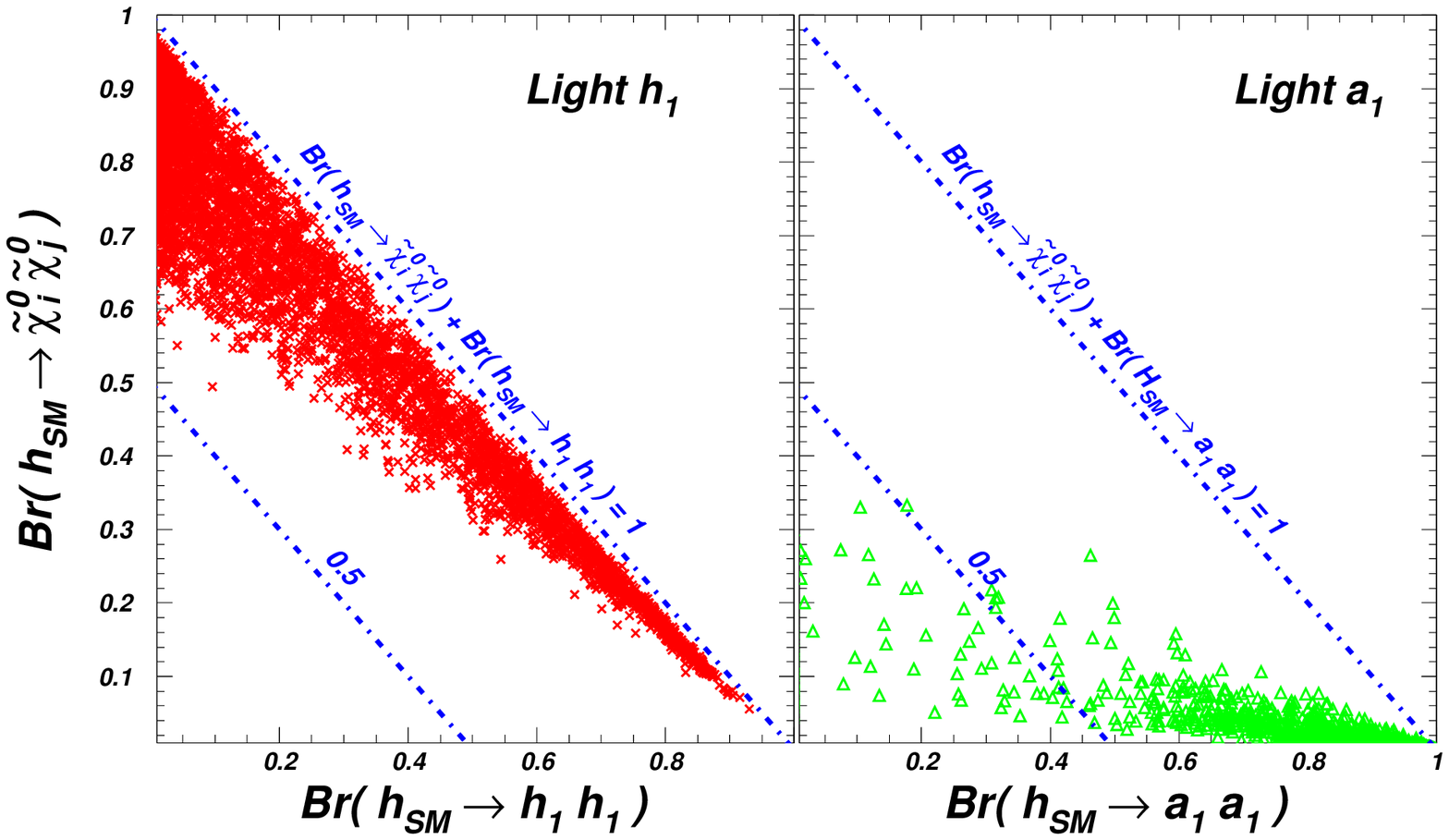} \hspace*{0.2cm}
\includegraphics[width=13.7cm]{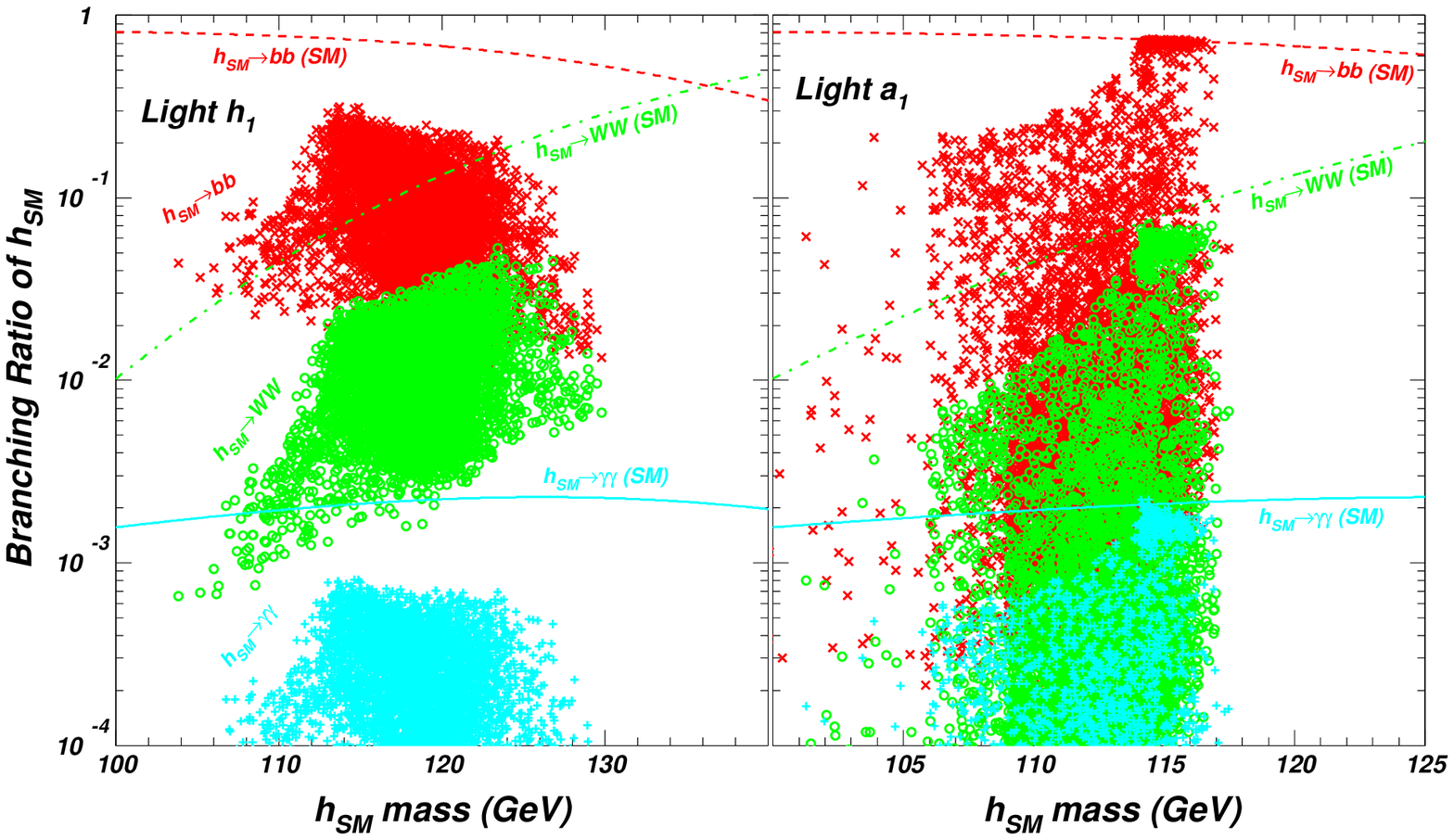}
\vspace{-0.8cm} \caption{Same as Fig.\ref{fig1}, but showing the
decay branching ratios of the SM-like Higgs boson $h_{\rm SM}$. Here
$Br(h_{\rm SM} \to \tilde{\chi}_i^0 \tilde{\chi}_j^0)$ denotes  the
total rates for all possible $h_{\rm SM} \to \tilde{\chi}_i^0
\tilde{\chi}_j^0$ decays.} \label{fig2}
\end{figure}
%%%%%%%%%%%%%%%%%%%%%%%%%%%%%%%%%%%%%%%%%%%%%%%%%%%%%%%

In  the light $\tilde{\chi}_1^0$ scenario, $h_{\rm SM}$ may decay
exotically into $\tilde{\chi}_i^0 \tilde{\chi}_j^0$, $h_1 h_1$ or
$a_1 a_1$, and consequently the conventional decays are reduced.
This feature is illustrated in Fig.~\ref{fig2}, which shows that the
sum of the exotic decay branching ratios may exceed $50\%$ and the
the traditional decays $h_{\rm SM} \to b \bar{b}, \tau \bar{\tau}, W
W^\ast, \gamma \gamma$ can be severely suppressed. Numerically, we
find that the branching ratio of $h_{\rm SM} \to b \bar{b}$ is
suppressed to be below $30\%$ for all the surviving samples in the
light-$h_1$ case and for about $96\%$ of the surviving samples in
the light-$a_1$ case. For the remaining $4\%$ samples in the
light-$a_1$ case, due to the kinematical forbiddance of the decay
$h_{\rm SM} \to a_1 a_1$, the ratio of $h_{\rm SM} \to b \bar{b}$
usually exceed $30\%$ and may even approach its SM value ($\sim
70\%$). The samples with the ratio exceeding $65\%$ are found to be
characterized by $m_{a_1}> 58 {\rm ~GeV}$, $m_{a_2} \ge 350 {\rm
~GeV}$ and $\tan \beta \ge 37$.

Another interesting feature  shown in Fig.~\ref{fig2} is that, due
to the open-up of the exotic decays,  $h_{\rm SM}$ may be
significantly lighter than the LEP bound. This situation is
favored by the fit of the precision electroweak data and is of
great theoretical interest \cite{Gunion}.

%%%%fig.3 %%%%%%%%%%%%%%%%%%%%%%%%%%%%%%%%%%%%%%%%%%%
\begin{figure}[htb]
\includegraphics[width=13cm]{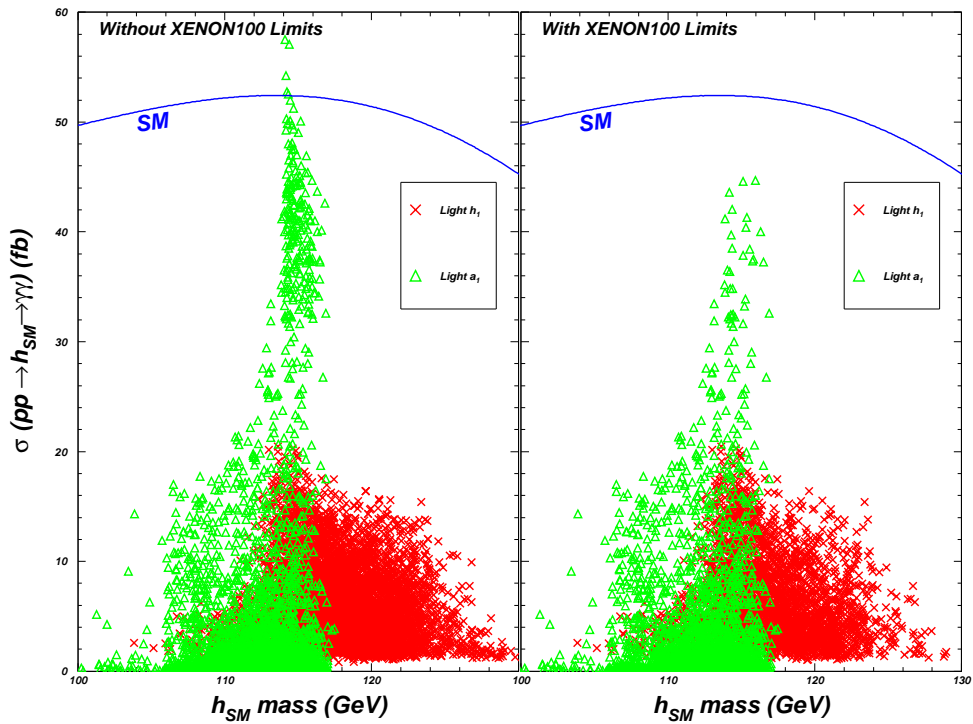}
\vspace{-0.8cm}
\caption{Same as Fig.~\ref{fig1}, but showing the
diphoton production rate of the SM-like Higgs boson at the LHC.}
\label{fig3}
\end{figure}
%%%%%%%%%%%%%%%%%%%%%%%%%%%%%%%%%%%%%%%%%%%%%%%%%%%%%%%

Since the conventional decay modes of $h_{\rm SM}$ may be greatly
suppressed, especially in the light-$h_1$ case which can give a
rather large $\chi-N$ scattering rate, the LHC search for
$h_{\rm SM}$ via the traditional
channels may become difficult. Noting $h_{\rm SM}$ is
bounded from above by about 130 GeV and hence its most important discovering
channel at the LHC is the di-photon signal, we show the di-photon
rate at the LHC with $\sqrt{s}=14$ TeV in Fig.~\ref{fig3}. In
calculating this rate, we used the narrow width approximation and
only considered the leading contributions to $p p \to h_{\rm SM}$ from
top quark, bottom quark and the squark loops.

Fig.~\ref{fig3} indicates that, compared with the SM prediction, the
NMSSM rate in the light $\tilde{\chi}_1^0$ scenario is suppressed to
be less than 20 fb for the light-$h_1$ case, and for the light-$a_1$
case most samples (about $96\%$) give the same conclusion. Since in
the light-$h_1$ case the $\chi-N$ scattering rate can reach the
CoGeNT sensitivity, this means that in the framework of NMSSM the
CoGeNT search for the light dark matter will be correlated with the
LHC search for the Higgs boson via the di-photon channel.

Note that, as shown in the left frame of Fig.\ref{fig3}, a few
samples can give a di-photon rate which is comparable with or
even exceeds its SM prediction. We checked
that these samples predict approximately same
decay branching ratios of $h_{SM}$ as the SM Higgs boson,
and the excess is mainly due to the slight suppression of
the width of $h_{SM} \to b \bar{b}$ so that $Br(h_{SM} \to \gamma \gamma)$
is enhanced. We also checked that the null result of the future XENON
6000 kg-days exposure will imply a Higgs di-photon signal below 20 fb
at the LHC with $99\%$ probability.

Finally, we point out that our light-$h_1$ scenario is different
from the scenario considered in \cite{Dark-Higgs}. The basic ideas
of \cite{Dark-Higgs} are: (1) Consider a special part in the
parameter space, which is characterized by $\lambda, \kappa \to 0$
and $A_\lambda \simeq \mu \tan \beta$ so that $\Det{\cal{M}}^2
\simeq 0$ to get a light $h_1$; (2) Consider a nearly decoupled
$\hat{S}$ so that the singlino serves as the dark matter with its
annihilation and its scattering with the nucleon proceeded mainly
by exchanging a light singlet $a_1$ and $h_1$ respectively. Such a
treatment obviously has the unnaturalness problem in electroweak
symmetry breaking since the condition $A_\lambda \simeq \mu \tan
\beta$ usually pushes the soft mass $m_{H_d}$ in
Eq.~(\ref{higgspot}) up to several TeV. We note the results of
\cite{Belanger} also suffer from this problem. In our scenario,
however, we keep the naturalness by requiring all soft masses to
be below TeV scale.

 {\em Conclusion:} We scrutinized the light neutralino dark matter
scenario in the NMSSM by scanning over the parameter space with all the
relevant soft masses below TeV scale. We found that in the parameter
space allowed by current experiments the neutralino dark matter
can be as light as a few
GeV and its scattering rate with the nucleon can reach the sensitivity
of XENON100 and CoGeNT (the CoGeNT signal can be explained).
The present XENON100 and CoGeNT data can exclude
a large parameter space, and the future 6000 kg-days exposure of
XENON100 can further explore (but cannot completely cover)
the remained parameter space.
In such a light dark matter scenario,  a light CP-even or CP-odd
Higgs boson must be present to satisfy the measured dark matter
relic density. As a result, the SM-like Higgs boson $h_{\rm SM}$ may
dominantly decay into a pair of light Higgs bosons or a pair of
neutralinos, and consequently the conventional decays like the
di-photon signal at the LHC will be much suppressed.
\vspace*{0.3cm}

{\em Acknowledgement} JMY thanks JSPS for the invitation
fellowship (S-11028) and the particle physics group of
Tohoku University for their hospitality.
This work was supported in part by NSFC
(Nos.~10821504,10725526,11075045,11005006), Doctor Foundation
of BJUT (No.~X0006015201102) from China and by the
Grant-in-Aid for Scientific Research (No.~14046201) from Japan.


\begin{thebibliography}{99}
\bibitem{Cao-bsg} We use the formula in \cite{NMSSMTools} to
 calculate $B \to X_s \gamma$.
\end{thebibliography}
\end{document}